\documentclass[preprint,prd,tightenlines,nofootinbib,superscriptaddress]{revtex4-1}

\usepackage{color}
\usepackage{graphicx}
\definecolor{red}{rgb}{1,0,0}
\def\lesssim{\ \hbox{\raise 2pt \hbox{$<$} \kern -13pt
                     \lower 3pt \hbox{$\sim$}}\ }
\def\greatersim{\ \hbox{\raise 2pt \hbox{$>$} \kern -13pt
                     \lower 3pt \hbox{$\sim$}}\ }

\def\cascade{{\sc Cascade}}
\def\pythia{{\sc Pythia}}

\def\powheg{{\sc Powheg}}
\input epsf.tex
\def\desepsf(#1 width #2){\epsfxsize=#2 \epsfbox{#1}}

\usepackage{amsmath,bm}
\begin{document}

\hspace*{11.5 cm} {\small DESY-11-262} 

\hspace*{11.5 cm} {\small IFJPAN-IV-2011-15}

\hspace*{11.5 cm} {\small OUTP-11-38-P}

\vspace*{1.4 cm} 

\title{Forward Jets  and  Energy Flow in Hadronic Collisions}
\author{M.\ Deak}
\affiliation{IFT-UAM/CSIC,  
Universidad Aut{\' o}noma de Madrid, E-28049 Madrid }
\affiliation{Departamento de  F{\' i}sica 
 de   Part{\' i}culas, Universidade de Santiago de Compostela,   
E-15782 Santiago de Compostela
 }
\author{F.\ Hautmann} 
\affiliation{Theoretical Physics Department, 
University of Oxford,    Oxford OX1 3NP}
\author{H.\ Jung}
\affiliation{Deutsches Elektronen Synchrotron, D-22603 Hamburg}
\affiliation{CERN, Physics Department, CH-1211 Geneva 23}
\affiliation{Elementaire Deeltjes Fysica, Universiteit Antwerpen, B 2020 Antwerpen}
\author{K.\ Kutak}
\affiliation{Instytut Fizyki Jadrowej im H. Niewodniczanskiego,  
 PL 31-342 Krakow }

\begin{abstract}
We observe that at the Large Hadron Collider, using forward + central  
detectors, it   becomes possible for the first time 
 to carry out calorimetric measurements of the  transverse  energy flow 
 due to ``minijets"  accompanying 
  production of   two jets separated by a  large rapidity interval.  
 We present  parton-shower  calculations of  energy 
 flow observables  in a  high-energy factorized     Monte Carlo   
 framework,    designed to take into account  
 QCD logarithmic corrections   both  in  the large rapidity  interval and  
 in    the  hard transverse momentum.    Considering  events with a forward and a  
 central   jet, we  examine   the   energy flow in the interjet region  and in the 
 region away from the jets. We   discuss the  role of these observables 
 to analyze multiple  parton collision  effects.  
\end{abstract}


\maketitle

The  production   of final states   created with high momentum transfers 
and  boosted to forward rapidities  
 is a  new  feature of the Large Hadron Collider 
compared to previous collider experiments,   subject of 
intense   experimental and theoretical  activity~\cite{ajaltouni}. 
Forward high-p$_\perp$ production enters the LHC  physics 
program in both new particle discovery processes (e.g.,  jet studies in 
decays of boosted massive states~\cite{boost1012})  and   
 new aspects of  standard model physics (e.g., 
 QCD at small $x$ and 
 its interplay with cosmic ray physics~\cite{ismd10}). 
 
Investigating  such final states  poses new challenges to both 
experiment and theory. On one hand, measurements of jet 
observables   in the   forward region   call for new experimental 
tools and analysis techniques~\cite{ajaltouni,denterria,hf-forwcal}. On 
the other hand, the  
evaluation of QCD theoretical  predictions  is made complex by the 
 forward kinematics   forcing    high-p$_\perp$  production 
into a region characterized by  multiple hard scales, possibly widely 
disparate from each other.   This raises the  issue  of  whether   
potentially large 
 corrections arise beyond finite-order perturbation theory  
 which call for  perturbative QCD resummations~\cite{muenav,hef,jhep09} 
 and/or contributions beyond single parton 
  interaction~\cite{sjozijl,bartfano,Sjostrand:2006za,pz_perugia,blok}.   
It is thus relevant to ask  to what extent current 
Monte Carlo   generators can provide realistic event 
simulations of forward particle production, and   how 
LHC experimental  measurements  can help  improve  
our understanding of  QCD effects in   the forward region.   

To this end,  in~\cite{jhep09,epr1012} we  have 
proposed  measuring  correlations  of a forward and a central jet  and 
performed a  numerical   analysis of the effects   
 of   noncollinear, high-energy 
  corrections to initial-state 
QCD showers.  First experimental studies have since appeared in 
preliminary form~\cite{cms-prelim}. The predictions~\cite{epr1012}  for the 
forward jet spectra are in reasonable agreement with the data.

In this paper we point out that the  
capabilities of forward + central detectors  at the LHC 
allow  one to perform  more detailed studies  of the event   structure,   
by measuring the associated  transverse 
energy flow  as a function of rapidity,   
both in the interjet  region and in the region away from the 
trigger dijets.    Such energy flow measurements have not been 
made before at hadron-hadron colliders. We observe that 
as a result of the   large phase space  opening up for    
high-p$_\perp$ production   at  LHC center-of-mass energies, 
one  has  an  average transverse energy flow 
per unit rapidity  of 10 GeV or more out to forward rapidity. Then it becomes 
possible to  carry out   measurements of  
the   flow  resulting from ``mini-jets"   with transverse energy 
above   a few GeV, thus suppressing the sensitivity of the 
observable   to  soft particle production.  
We  suggest this minijet energy flow as a way to investigate 
the detailed structure  of   events with forward and central jets. 

These  measurements   can be viewed as 
complementary  to  
measurements    performed  by  
  the CMS Collaboration~\cite{cms-pas-10-02}  
on the    energy  flow  in the forward direction  in minimum bias 
events and in events containing a central dijet system. 
The studies~\cite{cms-pas-10-02}  are 
designed to investigate properties of the soft underlying event; in particular, 
 they   illustrate  that the energy flow  observed in the forward region 
 is not well described by tunes of the \pythia\  Monte Carlo 
 generator~\cite{Sjostrand:2006za,pz_perugia} based 
 on charged particle spectra in the central region, especially for the 
 minimum bias sample. 
 The  energy flow   measurements discussed in this paper, on the other hand,  
   can serve to     investigate features of events  that depend on  (semi)hard   
   color radiation.    For proposed studies of  forward 
   event shapes  and correlations 
   to investigate minimum bias, see~\cite{skands11}.    

 Note that the   analysis  discussed  in this paper  can 
be extended to the case of 
  forward-backward jets.   Here one can   look   for   
Mueller-Navelet effects~\cite{ajaltouni,denterria,muenav}.   
Investigating  QCD radiation associated with 
  forward-backward jets will serve to analyze 
 backgrounds    in Higgs searches      from 
 vector boson fusion channels~\cite{vbf} and 
 studies based on a  central jet veto~\cite{ww-02} to extract information on Higgs 
 couplings~\cite{pilk}.   
In this case too   the underlying  jet activity accompanying the Higgs may receive 
comparable  contributions~\cite{deak_etal_higgs}  from  
  finite-angle radiative contributions to single-chain  showers,  
extending across  the whole  rapidity range, and from 
 multiple-parton interactions.

Our focus in this work is on initial-state effects    in  energy flow   measurements.   
Final-state effects such as those  discussed 
in~\cite{manch10,kucs}  may  require    resummation of logarithmically 
enhanced  terms  arising from restrictions on the phase space  in which  
color   radiation is considered and   depending  
on the  algorithms used to reconstruct the jets.  We do not address   these 
potential effects  in this paper.  
  See also~\cite{sung,hatta-ueda,bryan_eperp_10} 
  for  recent analyses of energy  flow observables.

We next discuss   the energy flow  
according to the method~\cite{jhep09,epr1012} and then 
 present  numerical  results.

\begin{figure}[htb]
\vspace{25mm}
\includegraphics{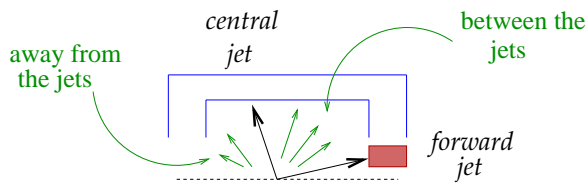}  
\caption{\it Production of forward and central jets: energy flow  
 in the inter-jet and outside 
regions. } 
\label{fig:jetcorr} 
\end{figure}

Let us  start by considering  final states associated 
  with the production of    
a forward and a central jet  at the LHC  as proposed  
in~\cite{epr1012}  (Fig.~\ref{fig:jetcorr}).   
To be specific, we take 
\begin{equation}
\label{rapkin}
 1 <  \eta_c  <  2  \;\;  ,  \;\;\;\;\;\;     
- 5  <  \eta_f  <  - 4       \;\;  ,   
\end{equation}
where  $\eta_c$   and  $\eta_f$ 
 are the central and forward jet   pseudorapidities.   
We consider the transverse energy flow as a function of  pseudorapidity 
\begin{equation}
\label{observ}
{ { d E_\perp } \over { d \eta}  }   =  { 1 \over \sigma}  \int dq_\perp \   q_\perp  \  
{ { d \sigma  } \over {dq_\perp \    d \eta}  }   \;\;  .  
\end{equation}
The  energy flow   is sensitive to color  radiation  associated with the 
trigger  specified  in Eq.~(\ref{rapkin}),  and its understanding  requires 
treating multiple QCD emission   by    parton showering or resummation 
methods beyond finite perturbative order.   

Observe that   the transverse  factor 
$q_\perp   $   in the  integrand on 
 the right hand side in Eq.~(\ref{observ}) 
enhances the sensitivity to the  high momentum transfer end 
of  the QCD parton  cascades compared to   the  
inclusive  jet  cross sections.  On one hand,     
   it  makes the transverse 
momentum ordering  approximation   less physically justified in the 
long-time  evolution  of the  parton  cascade.   On the other hand, 
 it   increases   the importance of    corrections 
 due to     extra hard-parton emission in the  
 jet production  subprocess at the  shortest  time scales. 
  
For these reasons, it is physically well-motivated to 
analyze the energy flow  by employing  the approach 
suggested in~\cite{epr1012}, in which  one couples 
the short distance  forward-jet  matrix elements~\cite{jhep09},  
which  contain extra hard-gluon emission 
via   high-energy factorization~\cite{hef}, 
  to the   transverse-momentum dependent 
   parton  showers~\cite{cascade_docu,jung02}, which go beyond  
the collinear  ordering approximation  by  Monte Carlo implementation of 
 CCFM evolution.\footnote{A  study  of  the  
  phenomenological   consequences of   the   dynamical 
effects at finite k$_\perp$,   in  both matrix elements 
and parton showers,  
  on final states with multiple jets  may be found in~\cite{hj_ang}.}   
 In this framework  we write the cross section in Eq.~(\ref{observ})  
as 
\begin{equation}
\label{forwxsec}
   {{d   \sigma  } \over 
{dq_\perp \    d \eta }} =  \sum_a  \int  \    \phi_{a/A}  \  \otimes \  
 {{d   {\widehat  \sigma}   } \over 
{  dq_\perp \    d \eta }}    \  \otimes \   
\phi_{g^*/B}    \;\; , 
\end{equation}
where  
$\otimes$ specifies  a convolution in both longitudinal and transverse momenta, 
$ {\widehat  \sigma} $  is the  hard scattering  cross section~\cite{jhep09},  
while  $ \phi_{a/A} $ 
and $ \phi_{g^*/B} $ are respectively  the collinearly 
factorized and high-energy factorized 
initial-state distributions, 
obtained 
as in~\cite{epr1012}   from,  respectively,    
the   on-shell and off-shell backward  shower evolution of the 
two    highly asymmetric incoming parton states    
that initiate the hard scatter.\footnote{See    Refs.~\cite{epr1012,hj_ang} for 
 discussions   of  the  use  of  the initial-state distributions in  
Eq.~(\ref{forwxsec}).} 
The approach~\cite{jhep09,epr1012}  
is designed to take into account  
 QCD logarithmic corrections   both  in  the large rapidity  interval and  
 in    the  hard transverse momentum.

Besides  the effects of  emission  of high transverse momenta,  
the energy flow (\ref{observ}) receives 
contributions from emission of  soft   particles.  
The   calculational framework   described above 
includes soft  parton emission  through  the 
$ 1 / (1 - z)  $ terms in  the branching  vertices  and 
the   form factors  
  in the shower  evolution~\cite{cascade_docu}, which contain 
  both   Sudakov and ``non-Sudakov"   contributions.   
We here point out 
 that,  unlike previous  collider experiments,  at the LHC it 
is possible to  reduce the  infrared sensitivity   of  
energy flow measurements  
by exploiting the large phase space available  
for production of high p$_\perp$,  as follows. 

Note that  
previous measurements of  transverse energy flow  were made  in 
lepton-proton collisions at HERA~\cite{h1-et}, where    one 
 had roughly an 
 average transverse energy of 1$\div$2 GeV per unit rapidity.   
 This increases by    a factor of five   at the LHC~\cite{cms-pas-10-02}     
  to about 5$\div$10 GeV or more per unit   rapidity, as a result of 
  the   phase space opening up for high-p$_\perp$ production. 
Besides the standard energy flow in 
  Eq.~(\ref{observ})  obtained by summing the energies over all  
 particles in the final states,    we will thus  consider    also  a  scenario in 
 which       first we merge particles into jets  by means of a  jet algorithm, 
 and then we construct   the associated energy flow from  jets  with transverse energy 
 above a given lower bound $q_0$, 
\begin{equation}
\label{qtbound}
 q_\perp >  q_0    \;\;   ,  \;\;\;\;\;\;   q_0 = {\cal O } ({\rm{a}} \;     {\rm{few}}  
 \;   {\rm{GeV}})  \;\;  .    
\end{equation}
The infrared safety is ensured by the use  of  the jet algorithm to cluster particles. 
On the other hand, 
the   use  of a lower  bound $q_0$ on 
 the order of a few GeV is made possible by the large transverse energy per unit 
 rapidity at the LHC.  In the calculations that follow 
we take    $q_0 = 5$  GeV   in   Eq.~(\ref{qtbound}). 
We call this   the   associated ``mini-jet" energy flow.   

In addition to  radiative  corrections from multiple  emission in a single 
parton collision, the  evaluation of $  { d E_\perp } /  { d \eta} $  is sensitive 
to    contributions of multiple parton 
collisions~\cite{bartfano}.  
 In the calculations that follow we will estimate  
these contributions using    the  Monte Carlo model~\cite{pz_perugia}.

In the following we consider  events with 
  dijets   (Fig.~\ref{fig:jetcorr})    
 reconstructed using  the  Siscone algorithm~\cite{fastjetpack}   with $R = 0.4$,         
in the rapidity region  (\ref{rapkin}). 
We will consider the cases   of jets with transverse energy  
E$_\perp\!>\! 10$~GeV.

\begin{figure}[htb]
\vspace{60mm}
\includegraphics{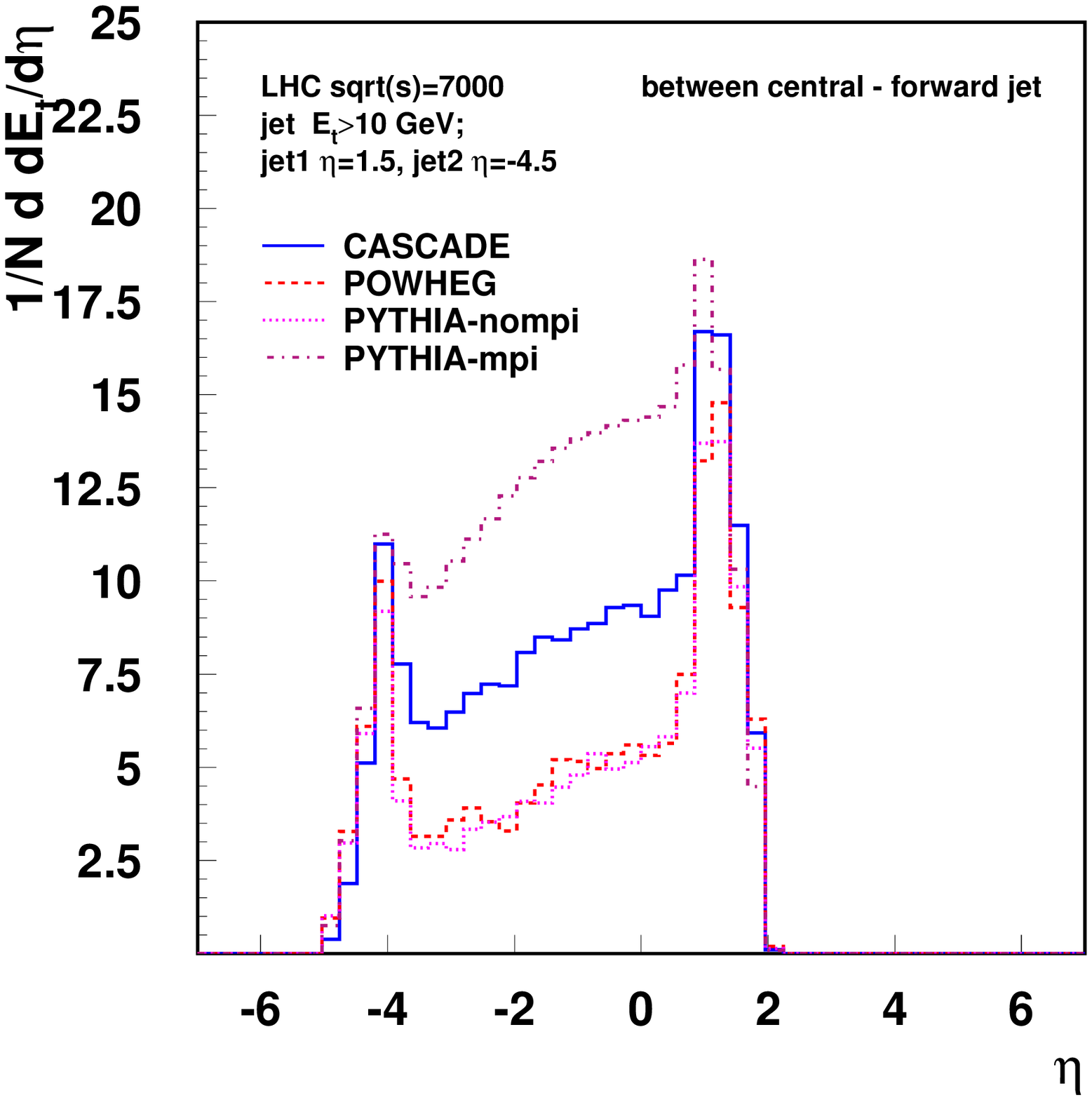}
\includegraphics{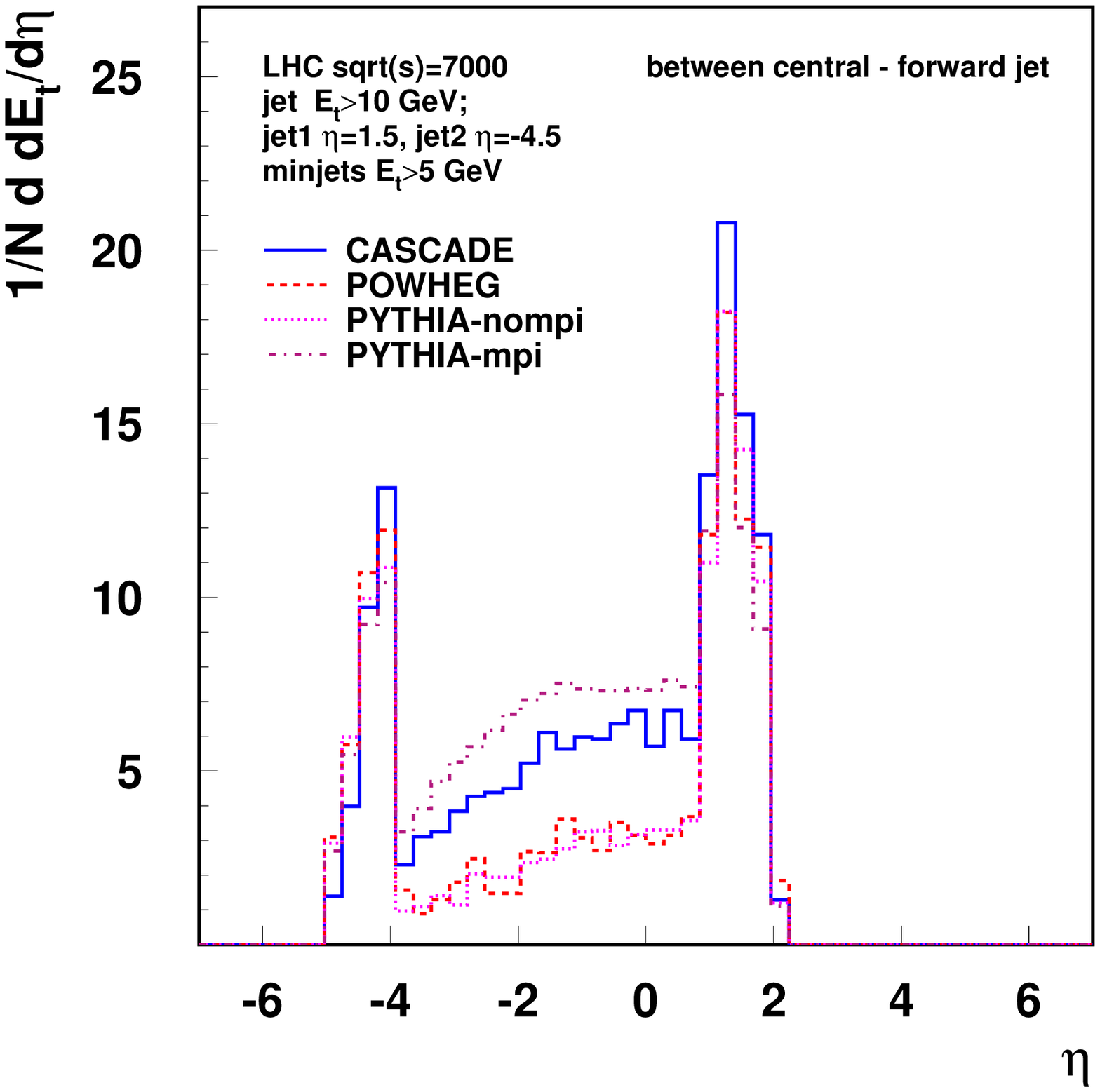}  
\caption{\it  Transverse 
energy flow in the  inter-jet   region:  (left) particle flow; (right) minijet flow. } 
\label{fig:betw} 
\end{figure}

Fig.~\ref{fig:betw}  shows the   transverse energy flow  in the interjet 
region for the cases of particle flow and of minijet flow. 
Besides the  calculation above given by the curves labelled \cascade, we 
report  results  obtained from   \pythia~\cite{pz_perugia} and \powheg~\cite{alioli}  
Monte Carlo event generators.  \pythia~\cite{pz_perugia} is used in two different modes, 
with  multiple parton interactions (\pythia-mpi, tune Z1~\cite{rickstune})  
and without  multiple parton interactions (\pythia-nompi).  
The particle energy flow   plot     on the left in Fig.~\ref{fig:betw}   
shows  the  jet profile picture,  and indicates     enhancements    of  
the energy flow   in the inter-jet region 
with respect to the \pythia-nompi    result 
 from higher order emissions in  \cascade\  and from multiple parton collisions in 
 \pythia-mpi. On the other hand, there is  little effect 
   from  the next-to-leading 
 hard correction  in \powheg\   with respect to  \pythia-nompi. 
  The energy flow  is dominated by
multiple-radiation, parton-shower effects.

 The minijet energy flow  plot   on the right in 
Fig.~\ref{fig:betw}   
  indicates similar 
effects,   with reduced sensitivity  to infrared radiation. 
As the minijet flow definition suppresses the
contribution of soft radiation,    the  \cascade\  and  \pythia-mpi results become 
more similar in the inter-jet  region. Particle and minijet flow measurements could be 
used to  investigate   the detailed form of 
multiparton  radiation effects. In particular, 
  these results  are  of interest for the 
QCD tuning of Monte Carlo generators,  especially   in connection 
with the estimation of QCD backgrounds in  search channels  
involving  two   jets far apart  in rapidity such as Higgs boson searches 
from vector boson fusion~\cite{pilk,vbf}. 

\begin{figure}[htb]
\vspace{60mm}
\includegraphics{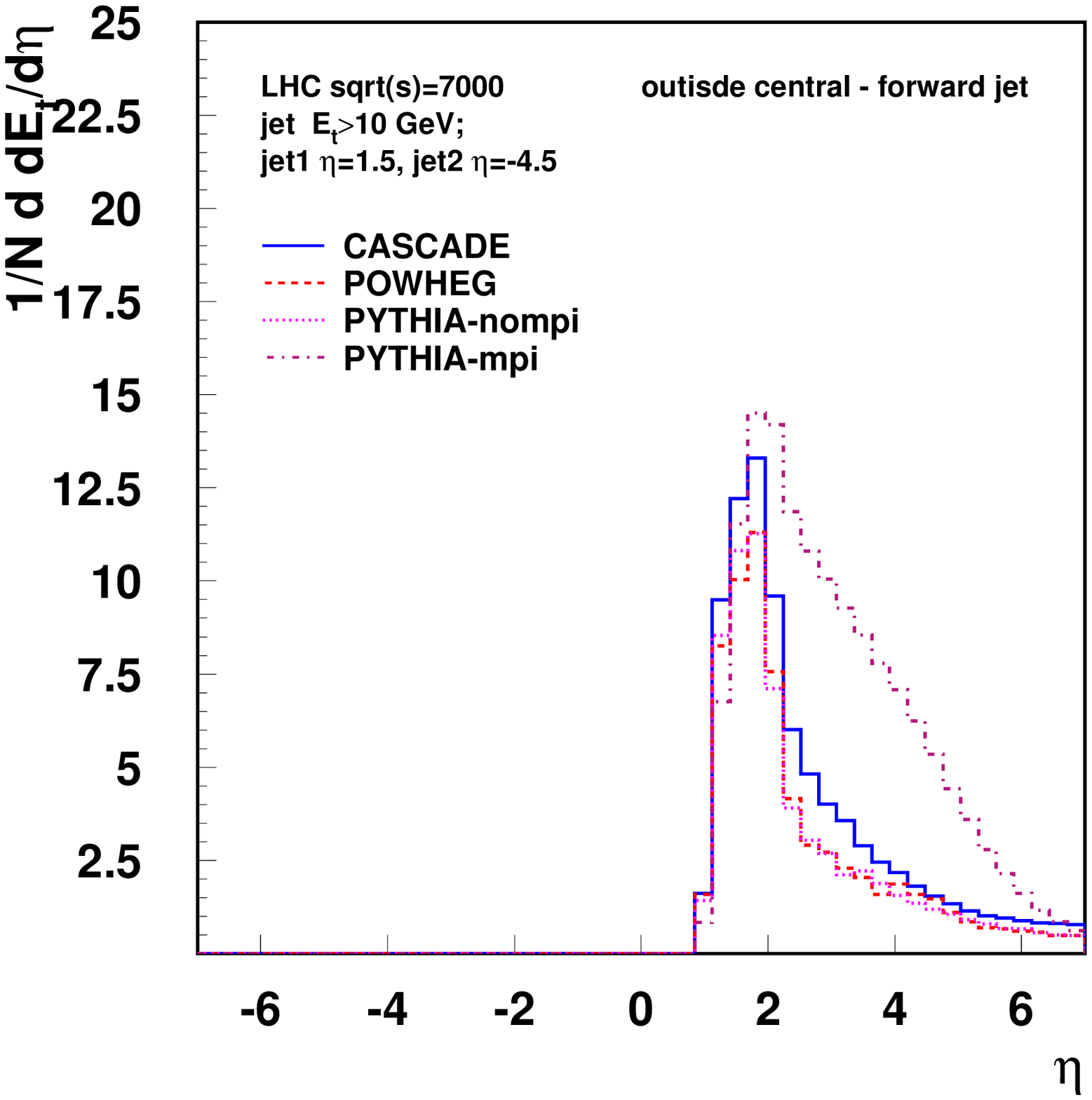}
\includegraphics{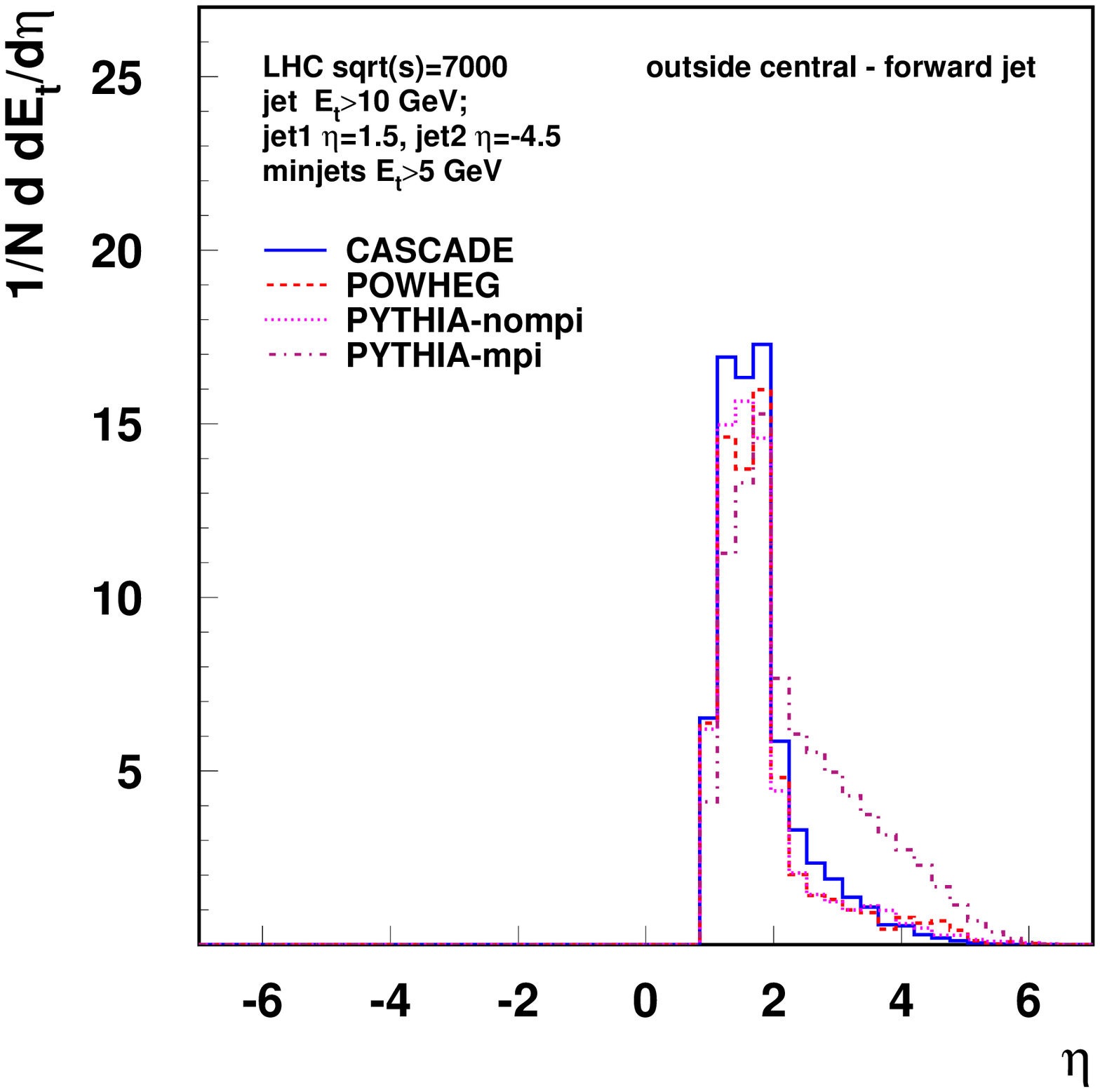}  
\caption{\it   Transverse  
energy flow in the outside region:  (left) particle flow; (right) minijet flow. 
  } 
\label{fig:outs} 
\end{figure}

It is   also   of interest   to measure the energy flow  in the  
  outside  region  corresponding to rapidities   opposite to the forward jet, 
far   in the backward  region. 
Results for this are shown in Fig.~\ref{fig:outs}.    
In the outside region an enhancement of  the 
 energy flow  can be produced by showering 
 from multiple parton  chains, while single-chain calculations give a 
 suppression of the transverse flow,  mainly due to phase space. 
In this region 
 one is sampling also  contributions from 
  the initial-state decay chain  at substantially larger 
 values of   longitudinal momenta,  where  
 the effects  of   corrections  to collinear ordering, taken into 
 account by the \cascade\ result,   are  not large. 
On the other hand,   contributions from
  multiple showers could be significant,  due to  
  gluon radiation  shifting   to larger  values of $x$  in each of the 
  sequential parton chains,    as  the total energy  
  available to the collision   is shared between   the  different   chains.

\begin{figure}[htb]
\vspace{53mm}
\includegraphics{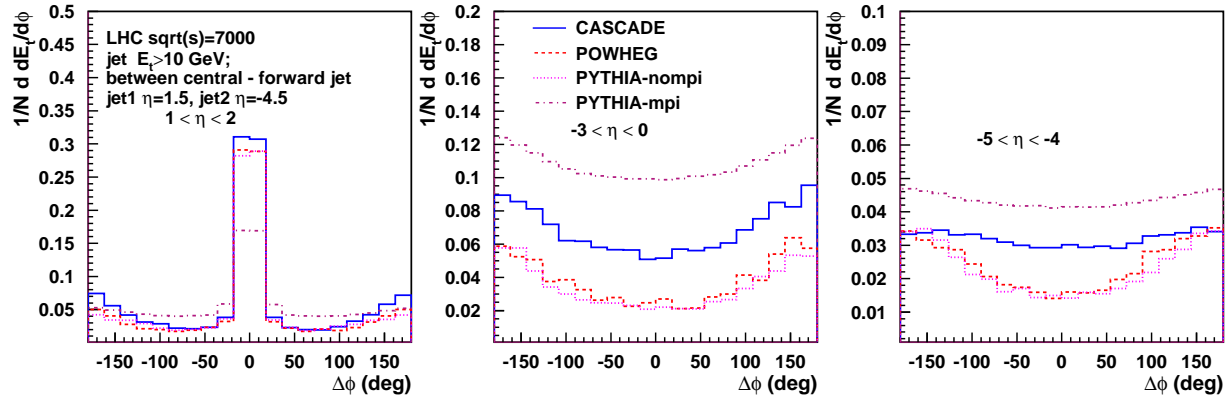}
\caption{\it  Azimuthal dependence of the  particle  
energy flow   for different rapidity ranges:   
(left)~central-jet;   (middle) intermediate;  (right) forward-jet. 
  } 
\label{fig:azimpart} 
\end{figure} 
 
\begin{figure}[htb]
\vspace{53mm}
\includegraphics{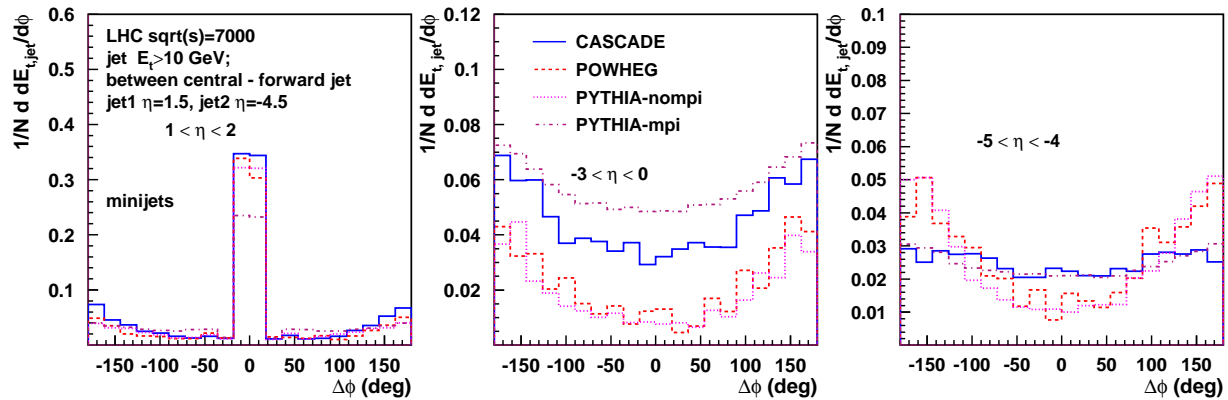}
\caption{\it  Azimuthal dependence of the  minijet   
energy flow   for different rapidity ranges:   
(left)  central-jet;   (middle) intermediate;  (right) forward-jet. 
  } 
\label{fig:azim} 
\end{figure}

 In Figs.~\ref{fig:azimpart} and \ref{fig:azim} we examine the azimuthal dependence 
of the transverse energy flow, respectively for the particle flow and minijet scenarios. 
Here $\Delta \phi$   is  measured  with respect to the central jet.   The 
$\Delta \phi$  distribution is shown for three different rapidity ranges, 
corresponding  to the central-jet,  forward-jet,    and intermediate  rapidities.

We see from  Figs.~\ref{fig:azimpart} and \ref{fig:azim} 
  that,  as we go  toward  forward rapidity,   the 
 \cascade\  and  \pythia-mpi calculations  give   a more   
 pronounced flattening  of the $\Delta \phi$  distribution compared 
 to  \powheg\   and   \pythia-nompi,    
  corresponding to   increased decorrelation between the jets. 
  The effect occurs  for  both the particle flow and the minijet flow. 
  In the minijet case (Fig.~\ref{fig:azim})    the  sensitivity to soft radiation is reduced. One 
  of the by-products of this is that also   the difference between  the 
    \cascade\  and  \pythia-mpi results      decreases.

Let us summarize. 
Forward jets  observed at the LHC for the first time 
  are   roughly   in agreement 
with  predictions from different Monte Carlo event generators. 
In this note we have proposed more exclusive measurements  
to probe the  associated event structure.  

We have studied the  transverse energy 
flow associated with the hadro-production of a forward and a central jet. 
These measurements can be performed with the 
LHC  by  exploiting the forward calorimeters;  
their  sensitivity to soft particle production  can be reduced   
 by  exploiting   the large phase space 
available for     high-p$_\perp$ production   at  LHC center-of-mass energies. 
The energy flow 
studies we have considered 
 can be helpful  for the QCD tuning of Monte Carlo event 
generators and for searches of new physics signals  based on  
analyses  of   jet  final states. 

We have presented Monte Carlo results that contain the 
small-$x$, large-rapidity resummation of QCD logarithmically 
enhanced terms based on the forward-jet matrix 
elements~\cite{jhep09,epr1012}. 
This is combined with large-$x$ effects according to the CCFM prescription 
implemented in the parton shower event 
generator~\cite{cascade_docu,jung02}. We have  
obtained results  
for the  energy flow  in the  inter-jet region and in the region away from the jets. 
We have  also estimated  the effects of  the multiple  parton interactions  
taken into account by~\cite{pz_perugia}, and found that  measurements of the 
energy flow in the inter-jet and outside regions can provide distinctive features 
to analyze such effects.

Phenomenological studies 
of color radiation activity 
along the lines suggested in this paper 
could be applied to  examining QCD backgrounds in  search channels 
from vector boson fusion~\cite{vbf,ww-02}. 
From  the theory  viewpoint,   
 energy flow observables  such as those considered here 
are  of  interest  both for studies of initial-state distributions that generalize 
ordinary parton distributions~\cite{hj_rec,becher-neub,xiao}    
 to   more exclusive 
  descriptions of    event   structure   and for studies of final-state 
 infrared effects~\cite{manch10,kucs,sung}  associated with  emission  of color 
 in restricted phase space regions. 

\vskip 1 cm

\noindent  
{\bf Acknowledgments}.   We thank the  
CERN Theory Division and  LHC Physics Center for hospitality.  
 M.~D.\  and F.~H.\   thank   UniverseNet  for   support. 
K.~K.\   acknowledges the support of NCBiR grant nr
LIDER/02/35/L-2/10/NCBiR/2011.

\end{document}